\documentclass[preprint]{vgtc}               

\ifpdf
  \pdfoutput=1\relax                   
  \usepackage{graphicx}                
  \DeclareGraphicsExtensions{.pdf,.png,.jpg,.jpeg} 
\else
  \usepackage{graphicx}                
  \DeclareGraphicsExtensions{.eps}     
\fi%

\graphicspath{{figures/}{pictures/}{images/}{./}} 

\usepackage{fontawesome5}
\usepackage{comment}
\usepackage{microtype}                 
\PassOptionsToPackage{warn}{textcomp}  
\usepackage{textcomp}                  
\usepackage{mathptmx}                  
\usepackage{times}                     
\usepackage{cite}                      
\usepackage{tabu}                      
\usepackage{booktabs}                  
\usepackage{hyperref}
\usepackage{url} 
\usepackage{enumitem} 

\onlineid{1041} 
\vgtccategory{Poster} 
\vgtcinsertpkg

\title{Do Vision-Language Models See Visualizations Like Humans?\\Alignment in Chart Categorization}

\author{
Péter Ferenc Gyarmati\thanks{e-mail: \href{mailto:peter.gyarmati@univie.ac.at}{peter.ferenc.gyarmati@univie.ac.at}}\\
    \scriptsize University of Vienna
\and
Manfred Klaffenböck\thanks{e-mail: \href{mailto:manfred.klaffenboeck@tuwien.ac.at}{klaffenboeck@cg.tuwien.ac.at}}\\
    \parbox{1.4in}{\scriptsize \centering TU Wien \\ University of Vienna}
\and
Laura Koesten\thanks{e-mail: \href{mailto:laura.koesten@mbzuai.ac.ae}{laura.koesten@mbzuai.ac.ae}}\\
    \parbox{1.5in}{\scriptsize \centering MBZUAI \\ Austrian Institute of Technology \\ University of Vienna}
\and
Torsten Möller\thanks{e-mail: \href{mailto:torsten.moeller@univie.ac.at}{torsten.moeller@univie.ac.at}}\\
    \scriptsize University of Vienna
}

\teaser{
  \centering
  \includegraphics[width=0.95\textwidth]{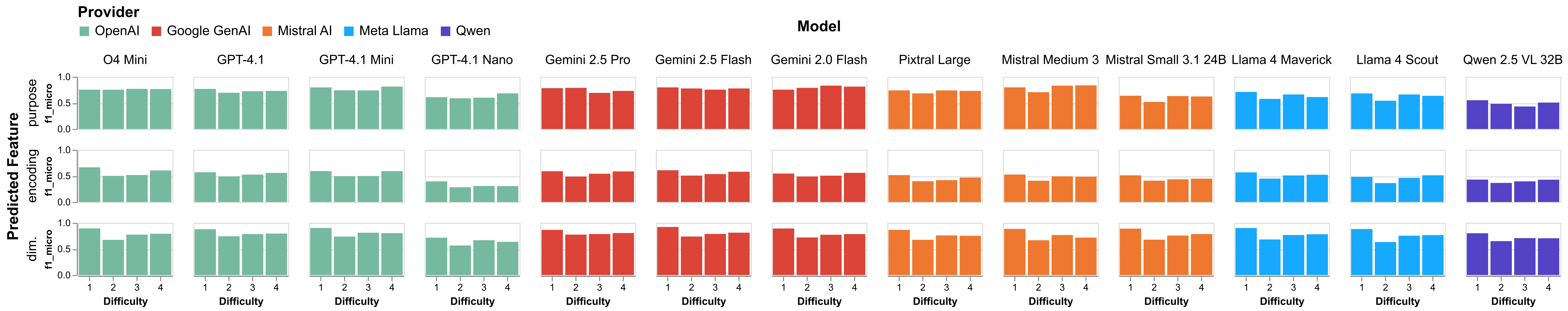}
  \caption{Micro-averaged F1-scores of Vision-language models (VLMs) in categorizing visualizations by \textsc{purpose}, \textsc{encoding}, and \textsc{dimensionality}, by human-assessed difficulty (1=easiest, 4=hardest). Higher bars mean better alignment with human expert perception. VLMs generally perform well on purpose and dimensionality but struggle with specific encoding types.}
  \label{fig:metric-overview-f1}
}

\abstract{Vision-language models (VLMs) hold promise for enhancing visualization tools, but effective human-AI collaboration hinges on a shared perceptual understanding of visual content. Prior studies assessed VLM visualization literacy through interpretive tasks, revealing an over-reliance on textual cues rather than genuine visual analysis \cite{Hong:2025:LLMLiteracy, Bendeck:2025:GPTEval}. Our study investigates a more foundational skill underpinning such literacy: the ability of VLMs to recognize a chart's core visual properties as humans do. We task 13 diverse VLMs with classifying scientific visualizations based solely on visual stimuli, according to three criteria: \textsc{purpose} (e.g., \textit{schematic}, \textit{GUI}, \textit{visualization}), \textsc{encoding} (e.g., \textit{bar}, \textit{point}, \textit{node-link}), and \textsc{dimensionality} (e.g., \textit{2D}, \textit{3D}). Using expert labels from the human-centric VisType typology~\cite{Chen:2025:Typology} as ground truth, we find that VLMs often identify \textsc{purpose} and \textsc{dimensionality} accurately but struggle with specific \textsc{encoding} types. Our preliminary results show that larger models do not always equate to superior performance and highlight the need for careful integration of VLMs in visualization tasks, with human supervision to ensure reliable outcomes.}

\begin{document}

\firstsection{Introduction}
\label{sec:introduction}
\maketitle

The integration of vision-language models (VLMs) into visualization tools promises to enhance data analysis \cite{Bendeck:2025:GPTEval}. However, effective human-AI collaboration depends on shared VLM-human perception. A perceptual gap can make VLM assistance unreliable or counterproductive. Measuring this gap is therefore a critical first step toward developing trustworthy AI-integrated visualization systems.

Existing research has evaluated VLM visualization literacy through analytical and interpretive tasks \cite{Bendeck:2025:GPTEval, Hong:2025:LLMLiteracy}, finding that VLMs often lean on textual prompts or pre-existing knowledge. We assess a more foundational skill: recognizing a chart's core visual properties, as human experts do. Before a model can \textit{interpret} a visualization's meaning, it must first accurately \textit{recognize} its fundamental attributes, such as its purpose, perceived dimensionality, and visual encoding. This foundational capability is critical for downstream applications. For instance, a tool for grammar-based editing cannot act on a command like "make the bars thicker" if it fails to recognize the "bars" in the first place, nor can a VLM in an agentic system narrate visual findings without first identifying the chart's components. Effective critique depends on the same prerequisite. We therefore explore how closely VLM perception aligns with that of human experts on this foundational parsing task, as a first step toward building more trustworthy and human-centered visualization tools.

Our study leverages two key resources. First, the VIS30K dataset \cite{Chen:2021:VIS30K} contains nearly 30,000 figures from 30 years of IEEE Visualization conference publications, offering a rich source of real-world scientific visualizations. Second, we leverage the VisType typology \cite{Chen:2025:Typology} as a human-annotated ground truth. This framework was developed by analyzing a subset of VIS30K, focusing on the \textit{essential stimuli} of each image. It provides expert labels for an image's primary \textsc{purpose} (e.g., a conceptual \textit{schematic}, a \textit{GUI} screenshot, or a data \textit{visualization}), its constituent visual \textsc{encoding} (e.g., \textit{bar}, \textit{line}, \textit{node-link}), its perceived \textsc{dimensionality} (e.g., \textit{2D}, \textit{3D}), and human-assessed difficulty of the categorization task itself.

Our experiment tests VLM perceptual alignment by restricting models to only two inputs: the raw figure image---which may include embedded text like axis labels, but excludes surrounding captions or body text---and the VisType category definitions provided via system prompt. This visual-only approach forces reliance on image analysis, allowing direct comparison with human experts and assessment of performance across models, scale, and human-perceived task difficulty.

\begin{figure*}[tbh!]
 \centering
 \includegraphics[width=0.975\textwidth]{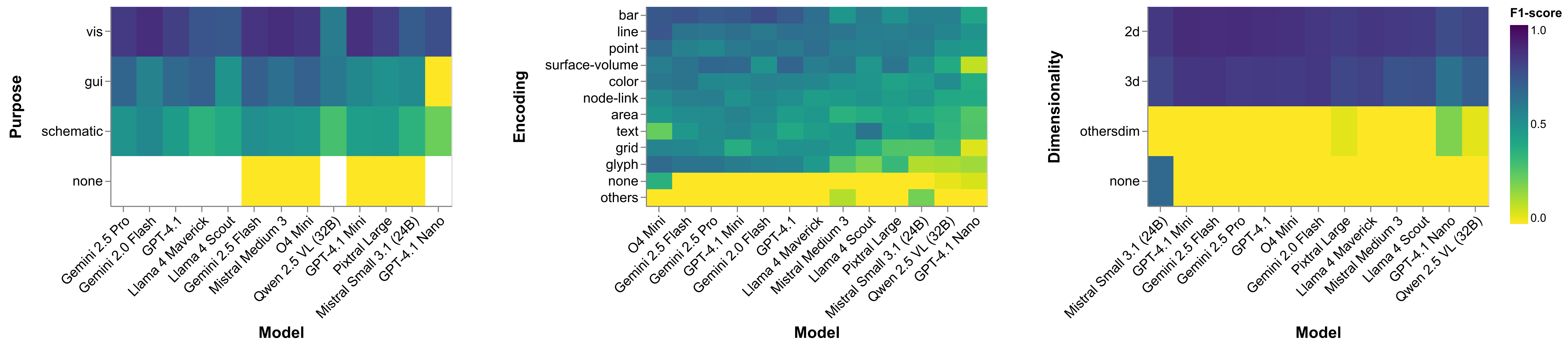} %
 \caption{F1-score heatmap comparing VLM classification to expert judgment for \textsc{purpose}, \textsc{encoding}, and \textsc{dimensionality}. Models (x-axis) and classification labels (y-axis) are sorted in descending order by their average F1-score. Darker colors mean higher agreement, yellow cells (F1-score $\approx 0$) indicate very low alignment, and white cells show where a model never produced the corresponding label at all. A \textit{none} label indicates cases where experts intentionally omitted a category. VLMs generally perform better at identifying \textit{2D} dimensionality, simpler encodings like \textit{bar}, \textit{line}, \textit{point}, and \textit{vis} (visualization example) purpose. However, classifying specific encoding types remains challenging. The prevalence of yellow and white cells reveals that models tend to offer an incorrect label rather than abstain, a behavior underscored by their difficulty predicting the \textit{none} category.}
 \label{fig:classification-accuracy-overview}
\end{figure*}

\section{Experimental Setup}
\label{sec:setup}
Our experiment has two phases: VLM inference on images and evaluation against expert labels.

\paragraph{Dataset and Models}
We use the dataset from Chen et al.'s VisImageNavigator application\footnote{Labeled dataset: \href{https://github.com/VisImageNavigator/VisImageNavigator.github.io/blob/54ab2319cca6a9e9056ce9cb5a337e920711b15e/public/dataset/vispubData30_updated_07112024.csv}{\faGithub~VisImageNavigator/VisImageNavigator.github.io}}, derived from VIS30K. The labels, based on the VisType typology~\cite{Chen:2025:Typology}, define images by \textsc{purpose}, \textsc{encoding}, \textsc{dimensionality}, and perceived \textsc{hardness}. A stratified sample of 305 images ensures representation across these categories. We test 13 diverse VLMs from five providers, spanning flagship and efficient models, as shown in \autoref{fig:metric-overview-f1}.

\paragraph{Inference and Evaluation Process}
Each VLM processes the 305 images in a zero-shot setting using default hyperparameters. The system prompt contains only the exact textual definitions for each category from Table 1 of Chen et al.'s VisType typology~\cite{Chen:2025:Typology}. We deliberately omitted visual examples to avoid priming the models' visual recognition. To handle structured output and uncertainty, we used models' native tool-calling to enforce a JSON schema. The schema definition for each category explicitly allowed a \textit{null} value, providing models a direct way to abstain when uncertain. For evaluation, we compare VLM outputs with the expert ground truth using standard multi-label classification metrics (e.g., F1-score). Code, prompts, and results are openly available\footnote{Code and data: \href{https://github.com/peter-gy/AutoVisType/}{\faGithub~peter-gy/AutoVisType}}.

\section{Preliminary Findings \& Discussion}
\label{sec:findings}
Our initial results (\autoref{fig:metric-overview-f1}--\ref{fig:classification-accuracy-overview}) reveal trends in VLM visualization perception. We report micro-averaged F1-scores, suitable for this multi-label task.

VLMs generally identify image \textsc{purpose} (e.g., \textit{vis}, \textit{schematic}) and \textsc{dimensionality} (especially clear \textit{2D}) with reasonable accuracy. However, discerning specific visual \textsc{encoding} types is a significant challenge. \autoref{fig:classification-accuracy-overview} (middle) shows that while simpler encodings like \textit{bar}, \textit{line}, and \textit{point} achieve moderate alignment, complex types like \textit{glyph}, \textit{grid}, or \textit{node--link}--and even \textit{color} or \textit{surface--volume}--are often misidentified. This suggests that VLMs, when relying solely on visual input without additional contextual guidance, struggle to capture the nuanced visual features human experts rely on when categorizing these encodings.

Notably, for this categorization task, larger models do not consistently outperform smaller ones. Some advanced models (e.g., Gemini 2.5 Pro, GPT-4.1) may better identify \textsc{purpose} but show no corresponding gain in recognizing specific \textsc{encodings}. This implies that for this task, model scale or recency alone does not guarantee better perceptual alignment.

Despite the option for a \textit{null} response for uncertainty, VLMs rarely used it, often assigning an incorrect category instead. This overconfidence is critical: a system misidentifying a line chart as a bar chart will offer nonsensical feedback, eroding trust. The \textit{none} category in \autoref{fig:classification-accuracy-overview} (where experts intentionally omitted a label) was also poorly predicted, underscoring this behavior. For instance, in identifying image \textsc{purpose}, 7 of the 13 tested models never produced a \textit{none} label, always forcing a different classification.

Human-assessed difficulty (\autoref{fig:metric-overview-f1}, x-axis) generally correlates with VLM performance: images that are harder to label for humans also tend to be harder for models. However, this correlation should not be mistaken for perceptual agreement. The models' poor performance on many charts labeled as 'easy' by human experts indicates that VLMs struggle with nuanced visual features. This reveals a perceptual gap: visual ambiguities that are easy-to-resolve for humans present challenges for VLMs.

\section{Conclusion \& Next Steps}
\label{sec:conclusion}

While modern VLMs make it easy to prompt a complex classifier from definitions, this accessibility masks a significant perceptual gap. Our findings show VLMs identify a chart's \textsc{purpose} and \textsc{dimensionality} but struggle with the fine-grained \textsc{encodings}, showing only partial alignment with expert perception. Model scale does not close this gap, and systems often guess incorrectly rather than express uncertainty. This cautions against blindly trusting VLMs, but also opens new design avenues for more robust human-AI collaboration. For instance, knowing which encodings a VLM perceives reliably can inform agentic systems that present a visualization to a user with an encoding optimized for human perception, while internally using a VLM-friendly encoding for more reliable automated analysis. Next, we will extend the benchmark to all labeled samples in the VIS30K corpus and inject controlled textual cues to probe truly multimodal reasoning. As VLMs evolve with greater performance and deeper multimodality, such benchmarks become crucial for tracking whether these advances translate to better human-AI alignment and trust. Through this open evaluation effort, we aim to guide both VLM development and visualization practice toward more transparent human-AI alignment in visualization systems.

\bibliographystyle{abbrv-doi}
\bibliography{template}

\end{document}